\documentclass[twocolumn,prl,floatfix,superscriptaddress]{revtex4-1}

\usepackage[linktocpage,bookmarksopen,bookmarksnumbered]{hyperref}
\usepackage{comment}
\usepackage{graphicx} 
\usepackage{gensymb}
\usepackage{amsmath,bm,amssymb}
\usepackage{amsfonts}
\usepackage[usenames]{color}

\begin{document}

\title{Induced magnetic two-dimensionality by hole doping in the superconducting infinite-layer nickelate Nd$_{1-x}$Sr$_x$NiO$_2$}

 \author{Siheon Ryee}
 \thanks{These three authors contributed equally to this work.}
 \affiliation{Department of Physics, KAIST, Daejeon 34141, Republic of Korea}

 \author{Hongkee Yoon}
 \thanks{These three authors contributed equally to this work.}
 \affiliation{Department of Physics, KAIST, Daejeon 34141, Republic of Korea}
 
 \author{Taek Jung Kim}
 \thanks{These three authors contributed equally to this work.}
 \affiliation{Department of Physics, KAIST, Daejeon 34141, Republic of Korea}
 
 \author{Min Yong Jeong}
 \affiliation{Department of Physics, KAIST, Daejeon 34141, Republic of Korea}
 
 \author{Myung Joon Han}
 \email{mj.han@kaist.ac.kr}
 \affiliation{Department of Physics, KAIST, Daejeon 34141, Republic of Korea}

\date{\today}

\begin{abstract}
	To understand the superconductivity recently discovered in Nd$_{0.8}$Sr$_{0.2}$NiO$_2$, we carried out LDA+DMFT (local density approximation plus dynamical mean-field theory) and magnetic force response calculations. The on-site correlation in Ni-$3d$ orbitals causes notable changes in the electronic structure. The calculated temperature-dependent susceptibility exhibits the Curie-Weiss behavior, indicating the localized character of its moment. From the low-frequency behavior of self-energy, we conclude that the undoped phase of this nickelate is Fermi-liquid-like contrary to cuprates. Interestingly, the estimated correlation strength by means of the inverse of quasiparticle weight is found to increase and then decrease as a function of hole concentration, forming a dome-like shape. Another finding is that magnetic interactions in this material become two-dimensional by hole doping. While the undoped NdNiO$_2$ has the sizable out-of-plane interaction, hole dopings strongly suppress it. This two-dimensionality is maximized at the hole concentration $\delta\approx0.25$. Further analysis as well as the implications of our findings are presented.
\end{abstract}
 
\maketitle

\section{Introduction}
Ever since its first discovery of high-temperature superconductivity in cuprates \cite{Bednorz}, tremendous theoretical and experimental  efforts have been devoted to the intriguing physics of this family of materials \cite{Keimer,PALee,Scalapino,Tsuei,Damascelli}. One important research direction is to find or design the cupratelike superconductivity in alternative transition-metal systems in the hope that scrutinizing the similarities and differences enables us to understand and utilize the exotic superconductivity \cite{Norman}. Although the unambiguous understanding of its nature has not been reached yet, some `essential' features for superconductivity have been highlighted. Obvious common features shared also with Fe-based superconductor include their two-dimensional electronic structure and magnetism \cite{Scalapino}. Spin $S=1/2$ moment interacting antiferromagnetically with each other, substantial $d$--$p$ hybridization, and the large orbital polarization (one-band physics) are of key importance in cuprates but not much relevant to Fe-based materials.

Nickelates have been studied along this line as a promising candidate of non-Cu-based but cupratelike superconductor \cite{Chaloupka,Uchida,Hansmann,MJHan,Boris,Zhang}. This long quest ends in a success by the discovery of hole-doped infinite-layer nickelate, Nd$_{1-x}$Sr$_{x}$NiO$_2$ grown on SrTiO$_3$ substrate \cite{Li}. While its mother compound, NdNiO$_2$, was originally synthesized more than a decade ago \cite{Hayward}, early theoretical studies excluded this form of nickelate from cuprate analogs due to the weak $d$--$p$ hybridization and noncupratelike Fermi surface \cite{KWLee}. In this regard, the recent discovery poses important challenges and possibilities for superconductivity research. Naturally, theoretical investigations intensively discuss its similarity with and the difference from cuprates in terms of electronic structure, gap symmetry and other details \cite{Botana,Sakakibara,Wu,Hepting,Jiang,Nomura,Gao}. While some important key features are identified to be similar with cuprates including the superconducting gap symmetry \cite{Sakakibara,Wu}, it is far from clear if the obvious differences are then become irrelevant or unimportant in this nickelate superconductor \cite{KWLee,Jiang}.

In this paper, we present several key findings obtained from first-principles calculations. (i) Our LDA+DMFT (local density approximation plus dynamical mean-field theory) calculation shows that the electronic correlation makes noticeable changes in the band structure and spectral weight. (ii) The calculated temperature-dependent local spin susceptibility shows that this material, albeit metallic, carries the local moment exhibiting Curie-Weiss behavior rather than Pauli-like. It is another notable similarity with cuprates. (iii) The undoped NdNiO$_2$ exhibits Fermi-liquid-like self-energy which is in contrast to lightly-doped as well as undoped cuprates. (iv) The correlation strength as estimated by the inverse of quasiparticle weight $Z^{-1}$ is enhanced and then reduced by hole doping, thereby forming a dome-like shape. (v) Magnetic force response calculations demonstrate that hole doping makes this system {\it magnetically} cupratelike. While the out-of-plane magnetic interactions are quite significant in NdNiO$_2$ contrary to the cuprate situation, they are largely suppressed by hole doping.
This {\it magnetic} two-dimensionality becomes maximized at around the doping concentration of the observed superconductivity.

\section{ Computation Methods}
We first carried out DFT (density functional theory) calculations within Perdew-Burke-Ernzerhof parametrization of generalized gradient approximation (GGA) \cite{PBE} as implemented in \textsc{VASP} \cite{vasp}. Following the literature \cite{Botana,Sakakibara}, we focused on LaNiO$_2$ instead of NdNiO$_2$ to avoid the numerical instabilities arising from Nd-$f$ electrons. To check the DFT magnetic ground state, four different magnetic orderings, namely ferromagnetic, A-type antiferromagnetic (AFM), C-type AFM, and G-type AFM orders, were compared using $\sqrt{2} \times \sqrt{2} \times 2$ unitcell. The $\mathbf{k}$-grid of $25 \times 25 \times 25$ was used with energy cutoff of 600 eV. To further check the possible structural effects on the magnetic ground state, we fully optimized the crystal structure with the force criterion of $10^{-3}$~meV/\AA. It is found that the G-type AFM order is the lowest energy state. The calculated total energy of FM, A- and C-type AFM is greater than G-type by 38, 38 and 5~meV/Ni, respectively, and the moment size is $M\simeq0.6$~$\mu_\mathrm{B}$. The results obtained from experimental lattice parameters of NdNiO$_2$ \cite{Li} will be presented as our main data. We also double checked our results with \textsc{OpenMX} software package \cite{openmx,Ozaki} and other exchange-correlation (XC) functionals including LDA \cite{PZ} and LDA/GGA$+U$ \cite{Anisimov}, confirming that our conclusion is valid. For LDA/GGA$+U$, we adopted the charge density formulations to avoid the ambiguities stemming from the spin density XC energy \cite{Ryee}. The on-site Coulomb interaction values of $U=F^0=3$, 5, and 7 eV were tested, and we consistently obtained the  G-type AFM order. It is found that, employing LDA gives the smaller magnetic moment ($M\simeq0.26$~$\mu_\mathrm{B}$/Ni) \cite{Ryee2}. 

We performed MFT (magnetic force theory) calculations \cite{Oguchi,Liechtenstein} to obtain magnetic exchange couplings on top of DFT electronic structures \cite{openmx,Ozaki,comment1}.
The calculations were carried out using our recently-developed MFT code, named \textsc{Jx} \cite{Yoon,Jx}. We adopted the most stable G-type AFM order obtained from GGA as the input. 
We also found that within the reasonable parameter range, LDA/GGA+$U$ gives the same conclusion.

The charge self-consistent LDA+DMFT calculations were performed by using \textsc{EDMFTF} package whose DFT part is based on \textsc{Wien2K} \cite{Haule}. {The muffin-tin radius $R_{MT}$ = 2.5, 1.98 and 1.7 {a.u.} for La,  Ni,  and O, respectively. The cut-off parameter to expand the basis set of $R_{MT}K_{max}$=7.0 was used.} Paramagnetic phase has been considered with the {\bf k}-grid of 11$\times$11$\times$14. We took the energy window of [$-$10, 10] eV to construct the local projector.
The Coulomb interaction and Hund's coupling of $U=F^0=5.0$ and $J_\mathrm{H}=(F^2+F^4)/14=0.8$ eV for Ni-$d$ states were adopted. Since this is a reasonable choice considering recent studies \cite{Sakakibara,Nomura}, the results of $U=5$ eV will be presented as our main data unless specified otherwise. We, however, have also performed calculations with larger $U$ values up to $U=9$ eV. Those results will also be presented below when necessary.
The DMFT impurity problem was solved via hybridization expansion CTQMC algorithm \cite{CTQMC1,CTQMC2}. The so-called nominal double-counting scheme was used \cite{nominal}. To simulate the doping, we used virtual crystal approximation within DFT-LDA. For the double-counting potential in LDA+DMFT under doping, the nominal charge of Ni-$d$ state was adjusted accordingly assuming that the doped charge resides on the Ni-site.


\begin{figure} [t] 
	\includegraphics[width=1.0\columnwidth, angle=0]{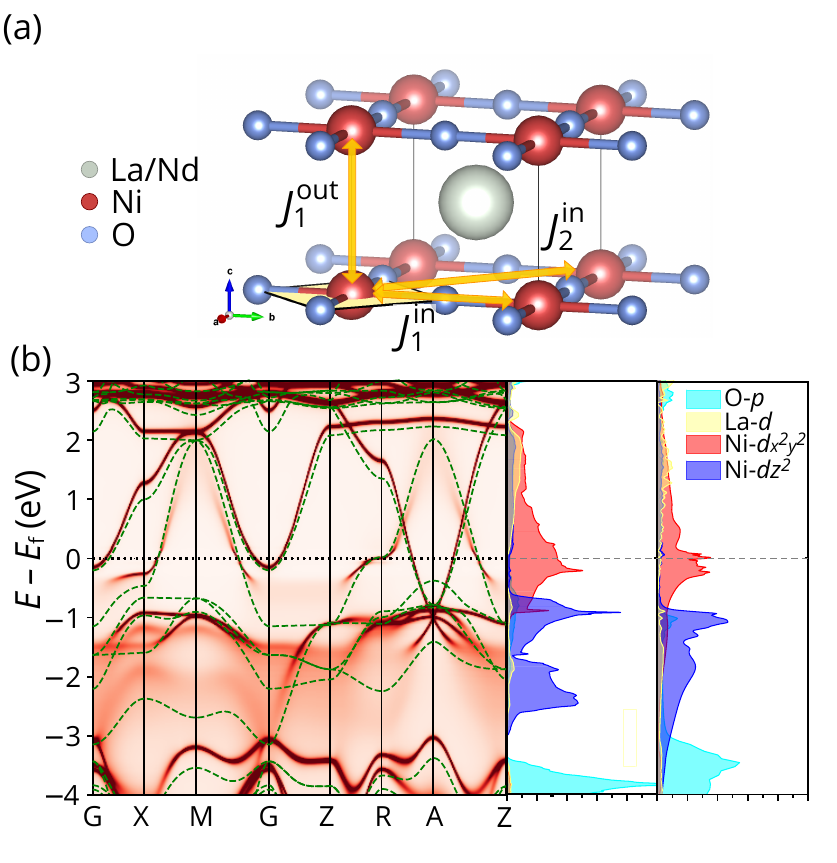}
	\caption{(a) Crystal structure of LaNiO$_2$ (which is identical with NdNiO$_2$ if La is replaced by Nd). Gray, red, and purple spheres represent La/Nd, Ni, and O atoms, respectively. The yellow arrows depict the first-neighbor out-of-plane ($J_1^\mathrm{out}$), the first-neighbor in-plane ($J_1^\mathrm{in}$), and the second-neighbor in-plane ($J_2^\mathrm{in}$) magnetic exchange couplings. (b) In the leftmost panel, the calculated band dispersion by LDA is presented in dashed green lines and the LDA+DMFT results are presented by color plot. Dark and bright red regions represent the larger and smaller spectral weights, respectively. The middle and right panel shows the calculated DOS by LDA (middle) and the spectral function by LDA+DMFT (rightmost).}
	\label{fig1}
\end{figure}

\section{Results}
\subsection{LDA+DMFT electronic structure} 
Having the realistic electronic structure and its analysis are the important first step toward further understanding of a given material. Due to the localized Ni-$3d$ orbitals, the correlation effect can be significant in this LaNiO$_2$. Previous band structure calculations, however, are limited either to LDA/GGA (which largely misses the effect of on-site correlations) or to LDA/GGA$+U$ (which is also quite limited, as a Hartree-Fock-like static approximation, in describing this metallic system). Here we adopt the LDA+DMFT method whose result is presented in Fig.~\ref{fig1}(b). In the left panel, the calculated band dispersion by LDA is presented in dashed green lines and the LDA+DMFT results are presented by color plot. Dark and bright red regions represent the larger and smaller spectral weights, respectively. It is clearly noted that the effect of correlation changes the electronic structure noticeably. In particular, $d_{x^2-y^2}$ bandwidth is markedly reduced near the Fermi level ($E_f$). See, for example, the states along $G ({\rm or}~ \Gamma)$--$X$--$M$ line and $G$--$Z$--$R$ line within the range of $E-E_f= -1$ to 0 eV. In spite of this correlation effect, however, the system remains metallic being consistent with experiment. We found that, up to the largest value of $U=9$ eV we tried, the metallic solution is obtained. The electron pockets centered at $G$ and $A$ have been highlighted as a distinctive feature of this material as it self-dopes the holes into Ni-$d$ bands. These pockets are also  maintained within LDA+DMFT. Upon hole doping, on the other hand, the electron pocket at $G$ gradually moves upward and eventually disappears from $E_f$, thereby forming the cupratelike one-band Fermi surface.

\subsection{The nature of magnetic moment} 
\begin{figure} [t] 
	\includegraphics[width=1.0\columnwidth, angle=0]{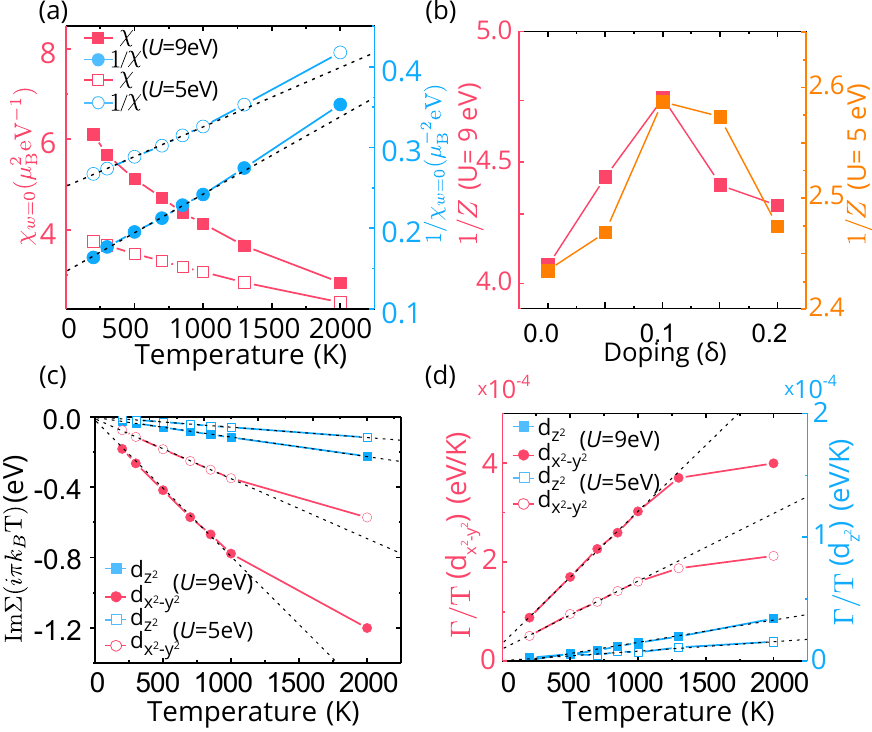}
	\caption{(a) The temperature-dependent local spin susceptibility $\chi_{\omega=0}$ (red) and the inverse susceptibility $\chi_{\omega=0}^{-1}$ (blue). The filled and open symbols represent the results of $U=9$ and 5 eV, respectively. (b) The calculated mass renormalization factor $Z^{-1}$ as a function of hole doping, $\delta$, by $U=5$ eV (orange) and 9 eV (red). (c) Imaginary part of self-energies at the lowest Matsubara frequency ($\pi k_B T$) as a function of temperature. The results of two $e_g$ orbitals are presented in different colors: $d_{z^2}$ (blue) and $d_{x^2-y^2}$ (red). (d) The calculated $\Gamma/T$ for $d_{z^2}$ (blue) and $d_{x^2-y^2}$ (red). See the main text for more details.}
	\label{fig2}
\end{figure}

Another fundamentally important question is about the nature of spin moments based on which further theoretical investigation and modeling can be developed in different ways. Figure~\ref{fig2}(a) presents our calculation results of temperature-dependent local spin susceptibility $\chi_{\omega=0} = \mu_\mathrm{B}^2g^2 \int_0^{1/(k_BT)} d\tau \langle S_z(\tau)S_z(0)\rangle$  (where $\tau$ is an imaginary time, $g$ is the spin gyromagnetic ratio, $k_BT$ is the temperature in eV, and $S_z$ is the local spin operator). Together with $\chi_{\omega=0}^{-1}$, it clearly indicates the Curie-Weiss behavior rather than Pauli-like. We note that $\chi_{\omega=0}$ at the larger $U$ value ($U=9$) shows the stronger local moment behavior and that the Curie-Weiss behavior is maintained even at $U=J_H=0$ (not shown). Thus it validates the various model studies that assume the local spin moments in this system.

\subsection{The nature of metallic phase} 
One obvious and physically important difference in between LaNiO$_2$ and cuprate is the undoped phase which is metallic only in the former. This fact immediately raises an important question regarding the nature of this chemically undoped metallic phase and its evolution as a function of hole concentration toward the superconductivity. In hole-doped cuprates, the undoped parent compound has a well-developed AFM order and insulating gap both of which are destroyed by doping. Importantly, before superconductivity arises, the intriguing pseudogap phase emerges which has long been a central topic of cuprate research \cite{pseudogap}. Further hole doping leads the system to be an exotic metallic phase, often called as `strange metal'. Fermi-liquid phase is eventually realized only in the heavy doping regime. Therefore, the question here is whether or not the undoped paramagnetic phase of LaNiO$_2$ is Fermi-liquid-like. and if not, whether its nature has any similarity with strange metal or pseudogap phase of cuprates.

In order to address this issue, we try to fit our calculated local self-energies at the lowest Matsubara frequency ($\pi k_B T$) onto the polynomials. The `first-Matsubara-frequency rule' states that the imaginary part of local self-energy at the lowest Matsubara frequency follows the $T$-linear scaling behavior in Fermi-liquid regime \cite{Chubukov,Hausoel}. 
We found that the calculated self-energies indeed exhibit the linear $T$-dependence for both $U=5$ and 9~eV (see Fig.~\ref{fig2}(c)). To further check the validity of this conclusion, we estimated the electron scattering rate $\Gamma=-Z\mathrm{Im}\Sigma(i\omega_n \to 0)$ (where $Z = \big(1-\frac{\partial \mathrm{Im}\Sigma(i\omega_n)}{\partial \omega_n} \big|_{\omega_n \to 0}\big)^{-1}$ is the quasiparticle weight and $\Sigma(i\omega_n)$ is the local self-energy on Matsubara frequency axis) by extrapolating the imaginary part of the Matsubara self-energy (Fig.~\ref{fig2}(d)). It is clearly seen that the estimated $\Gamma/T$ of Ni-$d_{x^2-y^2}$ and $d_{z^2}$ exhibits the linear $T$-dependence below $\sim 1000$~K, also indicating the Fermi liquid.
Hereby, we conclude that the undoped LaNiO$_2$ is Fermi-liquid-like, which is in a sharp contrast to cuprates. This intriguing undoped phase (carrying the local moment feature and simultaneously the Fermi-liquid nature) deserves further study.

\begin{figure} [t] 
	\includegraphics[width=1.0\columnwidth, angle=0]{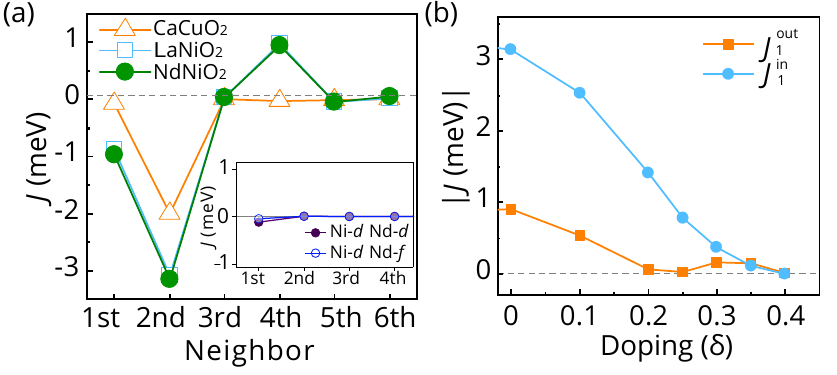}
	\caption{(a) The calculated $J$ profile as a function of Ni--Ni (or Cu--Cu) interatomic distance for LaNiO$_2$ (blue open squares), NdNiO$_2$ (green filled circles; with $U$=8 eV for Nd-4$f$), and CaCuO$_2$ (open triangle; orange). The first, second, and fourth neighbor coupling corresponds to $J_1^\mathrm{out}$, $J_1^\mathrm{in}$, and $J_2^\mathrm{in}$, respectively (see also Fig.~\ref{fig1}(a)). The inset shows that the calculated $J$ in between Ni-3$d$ and Nd-5$d$ (filled purple circles), and Ni-3$d$ and Nd-4$f$ (open blue circles) are both negligibly small. The negative and positive sign of $J$ represents the AFM and ferromagnetic couplings, respectively. (b) The calculated $J_1^\mathrm{out}$ and $J_1^\mathrm{in}$ as a function of hole doping $\delta$.  For comparison, their absolute values are presented.}
	\label{fig3}
\end{figure}

\subsection{The magnetic interactions}
Magnetic interactions have been considered as the key to understand the unconventional superconductivity including not just cuprate family but also Fe-based and heavy fermion materials. The estimation of these couplings is therefore a crucial step. Here we performed the MFT calculations to obtain the magnetic coupling constant $J$'s in between Ni sites. The results are summarized in Fig.~\ref{fig3}(a). The first thing to be noted is that LaNiO$_2$ has a sizable out-of-plane coupling. The first neighbor out-of-plane coupling, $J^{\rm out}_1$, is indeed the nearest neighbor interaction among Ni sites (see Fig.~\ref{fig1}(a)). Notably, this interaction is about 29\% of the strongest interaction $J^{\rm in}_1$ which refers to the in-plane nearest-neighbor coupling (the second nearest neighbor overall). 
This is another striking difference from cuprate.
In cuprates, it is well known that the magnetic interactions are basically two-dimensional and the out-of-plane coupling is negligible. In fact, our calculation result of isostructural infinite-layer cuprate compound confirms it: As shown in Fig.~\ref{fig3}(a), $J^{\rm out}_1$ for CaCuO$_2$ is very small. It is therefore tempting to conclude that, if the magnetic interaction is responsible for superconductivity in any sense, the newly discovered nickelate is the different case from cuprate superconductivity.

Remarkably, however, this three-dimensional nature of magnetic interaction profile is changed to be two-dimensional by hole doping. As shown in Fig.~\ref{fig3}(b), hole dopings reduce the size of $J^{\rm out}_1$ significantly. At the hole doping of 0.2~--~0.25/f.u., it becomes basically zero. 
Note that the experimental value of hole doping for realizing superconductivity ($\delta\approx0.2$) is in good agreement with it. Although the nominal value of charge counting both in calculation and in experiment should not be considered as being identical, this excellent agreement is impressive and informative. Considering that the recent theoretical studies coincidentally report that the superconductivity in this nickelate is cupratelike in terms of gap symmetry for example \cite{Sakakibara,Wu}, our finding elucidates the effect of hole doping for inducing superconductivity: In this new nickelate superconductor, the main role of doping is neither to destroy the AFM order nor the insulating phase, but it is to make magnetic interaction be two-dimensional, and in this sense, cupratelike.

\begin{figure} [t] 
	\includegraphics[width=1.0\columnwidth, angle=0]{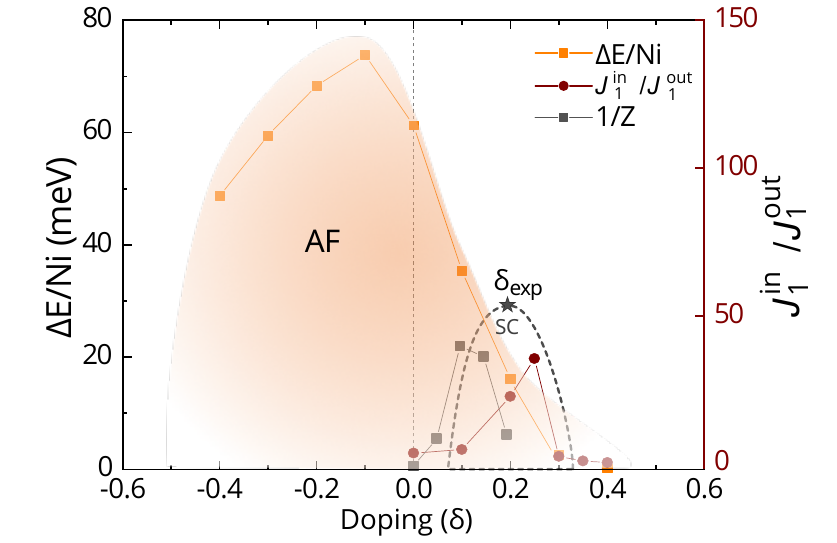}
\caption{A schematic summary of our calculation results showing the magnetic two-dimensionality and the correlation strength as a function of hole concentration. The orange squares represent the calculated $\Delta E$/Ni (where $\Delta E$ is the total energy difference in between the nonmagnetic and the lowest energy AFM order obtained from GGA). Brown circles present the two-dimensionality of the magnetic interaction measured by $J_1^\mathrm{in}/J_1^\mathrm{out}$. Black squares represent $Z^{-1}$ whose scale can be seen in Fig.~\ref{fig2}(b). For superconducting region, there is only one point experimentally known \cite{Li} at this moment (denoted by $\delta_{\rm exp}$). }
	\label{fig4}
\end{figure}

\section{ Discussion}
Figure~\ref{fig4} is a schematic summary of our results for magnetism and correlation. It is reminiscent of typical phase diagram of cuprate and many other high-T$_c$ materials. Here we also present the calculated total energy differences in between nonmagnetic and the lowest energy AFM order ($\Delta E= E^{\rm NM} - E^{\rm AFM}$); see the orange squares. Note that this $\Delta E$ represents the stability of AFM order (within DFT) and corresponds to the AFM ordering temperature. We also considered the electron doping region which is represented by negative $\delta$. Given that this system is `self-hole-doped', the behavior of AFM region as a function of $\delta$ is quite similar with other well-established phase diagram in both electron-doped and hole-doped side. It hopefully stimulates the further experimental and theoretical investigations of electron-doped side of this material.

In order to represent the two-dimensionality of magnetic interactions, we present the calculated value of $J^{\rm in}_1/J^{\rm out}_1$; see the brown circles in Fig.~\ref{fig4}. Its behavior as a function of $\delta$ forms also a domelike shape, first increases and then decreases. As noted above, it reaches its maximum at $\delta\approx$ 0.2~--~0.25 which is well compared with the experimental value. It suggests a possibility that the  magnetic two-dimensionality is essential for superconductivity.

Another interesting feature is the behavior of correlation strength as a function of $\delta$; see the black squares in Fig.~\ref{fig4}. As mentioned above, $Z^{-1}$ exhibits a domelike feature although it varies at most $\sim 0.2$ in our calculation of $U=5$ eV (with $U=9$ eV, the variation becomes slightly larger $\sim 0.7$); see Fig.~\ref{fig2}(b). According to our LDA+DMFT calculation, it reaches its maximum at around $\delta=0.1$. Here the care needs to be paid when comparing $\delta$ in LDA+DMFT with that in DFT. In our LDA+DMFT calculations, the hole doping is simulated by virtual crystal approximation in DFT-LDA, and by adjusting the nominal value of double-counting in DMFT. Thus, the doped holes are basically constrained to reside in Ni-$3d$ bands in our LDA+DMFT scheme (although the self-consistency can redistribute them) whereas, in DFT, holes can be distributed over the other bands. Considering this point, it seems feasible that the maximum point of $Z^{-1}$ can even be closer to $\delta=0.2$. This feature can possibly have implications for superconducting mechanism.

\section{Summary}
We report several findings and discuss their implications for recently discovered superconductivity in Nd$_{0.8}$Sr$_{0.2}$NiO$_2$. First, we present the realistic electronic structure by including the effect of on-site correlation within LDA+DMFT. The calculation of temperature-dependent spin susceptibility clearly identifies the localized nature of spin moments. The estimated correlation strength by means of the inverse quasiparticle weight is found to be enhanced and then reduced by hole doping. Interestingly, its maximum point is fairly close to the experimental value of nominal hole concentration to induce superconductivity. Finally, our MFT calculations demonstrate that the key role of hole dopings is to make the system magnetically cupratelike. The out-of-plane magnetic coupling is largely suppressed by hole doping, and therefore the magnetic interactions become two-dimensional. This two-dimensionality also reaches its maximum at around $\delta=0.2$.

\section{Acknowledgements} 
S.R. is grateful to H. Sakakibara for helpful comments.
This work was supported by BK21plus program, Basic Science Research Program (2018R1A2B2005204), and Creative Materials Discovery Program through NRF (2018M3D1A1058754).



\bibliography{ref}

\begin{thebibliography}{43}
\bibitem [{com()}]{comment1}%
  \BibitemOpen
  \href@noop {} {\enquote {\bibinfo {title} {The current version of our
  \textsc{Jx} \cite{Jx} is compatible only with openmx for the option of full
  band calculations. therefore we used the electronic structure obtained by
  openmx for our mft calculations. we carefully checked that using different
  code does not make any noticeable difference in the electronic structure},}\
  }\BibitemShut {NoStop}%
\end{thebibliography}

\end{document}